%% file: main.tex
\documentclass[a4paper,onecolumn,11pt]{quantumarticle}
\pdfoutput=1
\usepackage[utf8]{inputenc}
\usepackage[english]{babel}
\usepackage[T1]{fontenc}
\usepackage{amsmath}
\usepackage{hyperref}
\usepackage{lineno,hyperref}
\usepackage{graphicx}
\usepackage{algorithm}
\usepackage[noend]{algpseudocode}
\usepackage{dcolumn}
\usepackage{bm}
\usepackage{braket}
\usepackage{amsmath}
\usepackage[table]{xcolor}
\usepackage{svg}
\usepackage[numbers,sort&compress]{natbib}
\usepackage{amsfonts}

\usepackage{tikz}
\usepackage{lipsum}

\begin{document}

\newcommand\independent{\protect\mathpalette{\protect\independenT}{\perp}}
\def\independenT#1#2{\mathrel{\rlap{$#1#2$}\mkern2mu{#1#2}}}

\title{Quantum Approximate Optimization Algorithm for Bayesian network structure learning}

\author{Vicente P. Soloviev}
\affiliation{Computational Intelligence Group (Universidad Politécnica de Madrid)}
\orcid{0000-0003-0050-0235}
\email{vicente.perez.soloviev@fi.upm.es}
\homepage{http://cig.fi.upm.es/CIGmembers/vicente\_perez}
\author{Concha Bielza}
\affiliation{Computational Intelligence Group (Universidad Politécnica de Madrid)}
\homepage{http://cig.fi.upm.es/CIGmembers/concha\_bielza}
\orcid{0000-0001-7109-2668}
\email{concha.bielza@fi.upm.es}
\author{Pedro Larrañaga}
\email{pedro.larranaga@fi.upm.es}
\orcid{0000-0003-0652-9872}
\affiliation{Computational Intelligence Group (Universidad Politécnica de Madrid)}
\homepage{http://cig.fi.upm.es/CIGmembers/pedro-larranaga}
\maketitle

\begin{abstract}
Bayesian network structure learning is an NP-hard problem that has been faced by a number of traditional approaches in recent decades. Currently, quantum technologies offer a wide range of advantages that can be exploited to solve optimization tasks that cannot be addressed in an efficient way when utilizing classic computing approaches. In this work, a specific type of variational quantum algorithm, the quantum approximate optimization algorithm, was used to solve the Bayesian network structure learning problem, by employing $3n(n-1)/2$ qubits, where $n$ is the number of nodes in the Bayesian network to be learned. 
Our results showed that the quantum approximate optimization algorithm approach offers competitive results with state-of-the-art methods and quantitative resilience to quantum noise.
The approach was applied to a cancer benchmark problem, and the results justified the use of variational quantum algorithms for solving the Bayesian network structure learning problem.

\end{abstract}

\section{\label{sec:introduction}Introduction}

Bayesian networks (BNs) are a family of probabilistic graphical models that compactly represent the joint probability distribution of a set of random variables \cite{koller2009probabilistic}. Some of the most important characteristics of this type of model are its capability of graphically representing the uncertain knowledge contained in data and the possibility of including expert knowledge in the model. For these reasons BNs are widely used in machine learning \cite{murphy2012machine} for different applications \cite{bielza2014bayesian, 9387117}.

Regarding the high computational demands associated with BNs, two main problems have been studied in the literature: inference, which involves calculating a posterior probability distribution for some system variables when observing the values of other variables; and structure learning, which involves finding the optimal BN graph that best fits some given data. This paper is focused on the latter type of problem.

The Bayesian network structure learning (BNSL) problem is known to be NP-hard \cite{chickering1996learning} because the number of possible structures for a Bayesian network with $n$ nodes $h(n)$ increases more than exponentially with the number of variables $n$ in the given data \cite{robinson1977counting}:
\begin{align}
    \label{possible_structs}
    h(1) &= 1 \nonumber\\
    h(n) &= \sum_{i=1}^{n}(-1)^{i+1} \binom{n}{i} 2^{i(n-i)} h(n-i),
\end{align}
and thus, heuristic search algorithms are commonly used. In classic computing a wide range of approaches, such as particle swarm \cite{aouay2013particle, quesada2021structure}, evolutionary algorithms \cite{blanco2003learning, larranaga1996structure}, simulated annealing \cite{lee2019parallel}, and tabu search \cite{tabu} have been applied to solve the BNSL problem in recent decades \cite{scanagatta2019survey}.

More recently, the capabilities of quantum computers to reduce the required execution time when facing different optimization tasks and to solve very complex problems that may not be approachable with classic computing methods have attracted much interest. Quantum computing \cite{nielsen2002quantum} is based on quantum mechanics principles such as quantum entanglement and quantum superposition, which allow quantum algorithms to explore areas of the search spaces of optimization problems in a parallel and more efficient way. 

Quantum annealing (QA) \cite{hauke2020perspectives} is a quantum heuristic that can solve certain optimization problems exponentially faster than classic approaches. The BNSL problem has been mapped to a quadratic unconstrained optimization problem (QUBO) to be solved by using QA \cite{o2015bayesian, shikuri2020efficient}.

In recent years, quantum machine learning (QML) \cite{schuld2018supervised} has attracted much attention. QML combines machine learning algorithms and quantum computing theory to construct new hybrid algorithms that exploit the benefits of both fields. An example of such a hybrid algorithm is a variational quantum algorithm (VQA) \cite{mcclean2016theory}, which is composed of a classic optimization loop that embeds a quantum subroutine. This routine measures a quantum parametric circuit (variational ansatz), while the classic loop optimizes the parameters of the quantum circuit in each iteration of the algorithm used to minimize the cost function. VQAs, such as the variational quantum eigensolver \cite{peruzzo2014variational} and the quantum approximate optimization algorithm (QAOA) \cite{farhi2014quantum}, are widely used for different combinatorial optimization problems \cite{behera2020solving, choi2019tutorial, shaydulin2019evaluating, fontana2021evaluating, verdon2017quantum}. 
Some studies \cite{streif2019comparison} have proven that some types of optimization problems have landscape dispositions that makes the quantum and simulated annealing methods converge to local optimal solutions, while the QAOA is able to overcome this limitation and provide better solutions. 
Quantum and simulated annealing have already been applied to BNSL; however, to the best of our knowledge, the use of the QAOA has not been found in the literature.
In this work, we address the BNSL problem with the QAOA, and analyse the performance of different variants of the algorithm. 

Currently, the state of the art of quantum computers is the noisy intermediate-scale quantum (NISQ) era, which is characterized by quantum computers with hundreds of qubits and no error correction. Thus, there is a need to develop algorithms that do not require a large number of qubits and that offer resilience to the presence of quantum noise (which characterizes quantum devices). VQAs, and QAOAs in particular, are some of the most promising algorithms in the NISQ era, as their implementations optimize the number of utilized qubits, and moreover, the variational ansatzs are expected to offer resilience to quantum noise such as amplitude and phase damping errors \cite{nielsen2002quantum}.
We also analyse the resilience of the algorithm to the presence of different types of quantum noise, in the particular case of the BNSL problem.


The paper is organized as follows. 
Section~\ref{sec:bn} describes the fundamental basis of BNs, the classic methods for learning the structures of these models, and the QUBO formulation in which this work is inspired.
Section~\ref{sec:qaoa} provides an overview of the QAOA approach. Section~\ref{sec:implementation} describes how the QAOA ansatz is built and the characteristics integrated in our approach.
Section~\ref{sec:results} analyses the performance of the QAOA approach, the resilience of the algorithm to quantum noise, and a real application of the algorithm for solving the BNSL problem. 
Finally, Section~\ref{conclusions} rounds the paper off with the conclusions of our work.

\section{Structure learning of Bayesian networks}
\label{sec:bn}

A BN can be properly defined as a pair $(G, \bm{\Theta})$ that represents a joint probability distribution over a set of random variables $\bm{X}=\{X_1, X_2, \dots, X_n\}$. Its representation is given by (i) a directed acyclic graph (DAG) $G = ({V, A})$, whose nodes $V$ correspond to the set of random variables, $X_1, \dots, X_n$, and arcs $A$ that represent the probabilistic dependencies among the variables; and (ii) a set $\bm{\Theta}$ of parameters that define the conditional probability distribution of any variable $X_i$ given its parents $\Pi_i$ in the graph, where the parents $\Pi_i$ of a variable $X_i$ are the nodes that have arcs that reach $X_i$.

Given this definition, the probability distribution $P(\bm{X})$ over the set of variables $\bm{X}$ is defined as the product of the conditional probability distributions of all variables:
\begin{equation}
    P(\bm{X}) = P(X_1, \dots, X_n) = P(X_1|\Pi_1) P(X_2|\Pi_2) \cdots P(X_n|\Pi_n) \nonumber
\end{equation}

The BNSL problem is a very complex NP-hard problem \cite{chickering1996learning} that is well-known in the state-of-the-art research due to the combinatorial explosion of possible DAGs which can represent the relationships among the variables in $\bm{X}$. Given a dataset $\mathcal{D}$ with $n$ columns and as many rows as variable observations, the objective is to determine the DAG that better reflects the relationships among the variables $\bm{X}=\{X_1, X_2, \dots, X_n\}$ found in $\mathcal{D}$.
Two variables $A$ and $B$ are said to be conditionally independent given $C$ if $P(A|B,C) = P(A|C)$, and thus, the values that $B$ takes contribute nothing to the certainty of $A$.
Three main BNSL approaches are available: (i) the score-based approach, whose objective is to optimize a function that evaluates the quality of the structure given the data; (ii) the constraint-based approach, which performs some statistical tests to check the conditional independences among the variables; and (iii) hybrid methods that combine both approaches. In this paper, we focus on implementing a score-based approach.


Some well-known scores have been used for the BNSL problem. The decomposability property is desirable for computational reasons. This means that the score of a structure given some data is computed as the sum of the local scores of the subgraphs formed by each variable $X_i$ and its parents $\Pi_i$,
$$
score(G, \mathcal{D}) = \sum_{i=1}^{n}score_{i}(\Pi_i, \mathcal{D}),
$$
where $G$ denotes a DAG. The objective of the optimization task is to maximize this score.

Relevant decomposable scores used for BNSL are the Bayesian information criterion (BIC) \cite{schwarz1978estimating}, K2 score \cite{cooper1992bayesian} and Bayesian Dirichlet equivalent uniform (BDeu) \cite{BDe}, among others.


\subsection{QUBO formulation of BNSL}
\label{sec:qubo}

In this section, we describe the original QUBO formulation introduced in \cite{o2015bayesian}, on which we base our approach. The formulation is based on four different Hamiltonians: $H_{score}$, which optimizes the likelihood of a structure given the input data; $H_{max}$, which ensures the maximum in-degree of each node to limit the Hamiltonian complexity; and $H_{trans}$ and $H_{consist}$ which guarantee that the adjacency matrix that represents the BN is acyclic. The computed QUBO expression is,
\begin{equation}
    \nonumber
    H(\bm{A}, \bm{R}, \bm{Y}) = H_{score}(\bm{A}) + H_{max}(\bm{A}, \bm{Y})
    + H_{trans}(\bm{R}) + H_{consist}(\bm{A}, \bm{R}),
\end{equation}
where $\bm{A}, \bm{R}$ and $\bm{Y}$ are the quantum bits associated to the adjacency matrix, topological order, and the maximum in-degree restriction variables, respectively, for variables $X_1, X_2, \dots, X_n$.

The QUBO problem formulation for solving a BNSL problem with $n$ nodes requires
\begin{equation}
    v_{size} = n(n-1) + \frac{n(n-1)}{2} + 2n
    \label{number_qubits}
\end{equation}
quantum bits, where $n(n-1)$, $\frac{n(n-1)}{2}$, and $2n$ are the number of bits associated with the adjacency matrix, the topological order, and the number of variables needed to restrict the maximum in-degree, respectively.

Each of the four different Hamiltonians that compose $H(\bm{A}, \bm{R}, \bm{Y})$ are deeply explained in this section.

\subsubsection{$H_{score}(\textit{\textbf{A}})$}

For $H_{score}$ we need to introduce the concept of an adjacency matrix ($\bm{A}$):
\begin{equation}
\label{eq_d}
    \bm{A} =
\begin{bmatrix}
a_{11} & \cdots & a_{1n} \\
\vdots & \ddots & \vdots \\
a_{n1} & \cdots & a_{nn}
\end{bmatrix}
\end{equation}
where $a_{ij} = 1$ if there exists an arc from $X_i$ to $X_j$ and $a_{ij} = 0$ otherwise. 

In this case, as BNs are represented as DAGs, the diagonal of this matrix is equal to zero, and thus, the bits of the diagonal are not required for the QUBO formulation. Then, $n(n-1)$ qubits are needed for the $H_{score}$ Hamiltonian to learn a BN of $n$ nodes, and
\begin{align}
H_{score}(\bm{A}) &= \sum_{i=1}^{n} H_{score}^{i}(\bm{a}_i) \label{h_score} \\
H_{score}^{i}(\bm{a}_i) &= \sum\limits_{\substack{J \subset \{1, \dots, n\} \setminus \{i\} \\ |J| \leq m}} (w_i(J) \prod_{j \in J}a_{ji}) \label{h_score_i}
\end{align}
where $\bm{a}_i = (a_{1i}, \dots, a_{ni})$, is the $i$-th column of $\bm{A}$, 
$w_i(J) = \sum^{\vert J \vert}_{l=0} (-1)^{\vert J \vert - l } \sum\limits_{\substack{K \subset J \\ \vert K \vert = l}} s_i(K)$, in which $s_i(K)$ is the score of node $i$ given the parent set $K$, and $m$ is the maximum in-degree allowed.

Note that the constant term is $w_i(\emptyset) = s_i(\emptyset)$, which refers to the score of node $X_i$ without its parents. If $X_i$ has a single parent $X_j$, then the above equation simplifies to
\begin{equation}
\nonumber
H_{score}^{i} (\bm{A}) = w_i(\emptyset) + w_i(\{j\})
= s_i(\emptyset) + s_i(\{X_j\}) - s_i(\emptyset)
= s_i(\{X_j\})
\end{equation}

Similarly, if $X_i$ has two parents $X_j$ and $X_k$,
\begin{equation}
\begin{split}
\nonumber
H_{score}^{i} (\bm{A}) & = w_i(\emptyset) + w_i(\{j\}) + w_i(\{k\}) + w_i(\{j, k\}) \\
& = s_i(\emptyset) + (s_i(\{X_j\}) - s_i(\emptyset)) + (s_i(\{X_j\}) - s_i(\emptyset)) + w_i(\{j, k\}) \\
& = s_i(\{X_j\}) + s_i(\{X_k\}) - s_i(\emptyset) + w_i(\{j, k\}) \\
& = s_i(\{X_j\}) + s_i(\{X_k\}) - s_i(\emptyset) + s_i(\{X_j, X_k\}) - s_i(\{X_j\}) - \\
& - s_i(\{X_k\}) + s_i(\emptyset)) = s_i(\{X_j,X_k\})
\end{split}
\end{equation}

\subsubsection{$H_{max}(\textit{\textbf{A}}, \textit{\textbf{Y}})$}

To ensure that the quantum algorithm only considers the maximum in-degree $m=2$ to restrict the search space, the $H_{max}$ Hamiltonian is implemented in a way such that $2n$ quantum bits are needed. These quantum bits are represented as a matrix:
$$
\bm{Y}=
\begin{bmatrix}
y_{11} & y_{12}\\
\vdots & \vdots \\
y_{n1} & y_{n2}
\end{bmatrix}
$$
where $y_{ij} \in \{0, 1\}$ are random binary variables. $\bm{Y}$ represents a slack variable used to reduce the inequality constraint of the maximum in-degree to an equality constraint.

The corresponding Hamiltonian results in $H_{max}(\bm{A}, \bm{Y}) = 0$ if the restriction is met, and $H_{max}(\bm{A}, \bm{Y}) > 0$ otherwise. Thus,
\begin{align}
    \nonumber
    H_{max}(\bm{A}, \bm{Y}) &= \sum_{i=1}^{n}H_{max}^{i}(\bm{a}_i, y_i) \\
    \nonumber
    H_{max}^{i}(\bm{a}_i, y_i) &= \delta_{max}(m - \sum_{j=1}^{n}{a_{ij}} - y_i)^2 \\
     &=
    \begin{cases}
      0, & d_i \leq m \\
      \delta_{max}(d_i - m)^{2}, & d_i > m
    \end{cases}  \nonumber
\end{align}
where $y_i = \sum_{l=1}^2 2^{l-1} y_{il}$, $i = (1, \dots, n)$, and $\delta_{max} \in \mathbb{R^{+}}$ is a prefixed penalization term.

\subsubsection{$H_{trans}(\textit{\textbf{R}})$ and $H_{consist}(\textit{\textbf{A}}, \textit{\textbf{R}})$}

To ensure the acyclicity of the adjacency matrix we need to implement two different Hamiltonians, $H_{trans}(\bm{R})$ and $H_{consist}(\bm{A}, \bm{R})$. The former uses the topological order to check the transitivity of the graph, and the latter checks the consistency between the topological order and the adjacency matrix.

A topological ordering of a directed graph is a linear ordering of its vertices such that for every arc $i \rightarrow j$ from vertex $i$ to vertex $j$, $i$ comes before $j$ in the ordering ($i<j$). The topological order is represented as
\begin{equation}
\nonumber
\bm{R}_{top} =
\begin{bmatrix}
r_{11} & \cdots & r_{1n} \\
\vdots & \ddots & \vdots \\
r_{n1} & \cdots & r_{nn}
\end{bmatrix}
\end{equation}
where $r_{ij}$ can be equal to $1$, only if $i \leq j$ and $r_{ij} = 0$ if $i > j$ for the given QUBO formulation. 
The lower triangular portion of $\bm{R}_{top}$ provides no additional information to the upper triangular portion of $\bm{R}_{top}$. 
Moreover, if a matrix is acyclic, then the trace of $\bm{R}_{top}$ is equal to zero. Considering this, the variables used for the QUBO formulation are represented as part of the matrix $\bm{R}_{top}$, where the diagonal of the matrix and all the elements below it have been removed
\begin{equation}
\label{eq_r}
\bm{R} =
\begin{bmatrix}
r_{12} & r_{13} & \cdots & r_{1n} \\
\ddots & r_{23} & \cdots & r_{2n} \\
\vdots & \ddots & \cdots & \vdots \\
\cdots & \cdots & \ddots  & r_{n(n-1)}
\end{bmatrix}
\end{equation}
Then $H_{trans}(\bm{R})$ is zero if the relation encoded in the $\bm{R}$ matrix is transitive and $\delta_{trans}$ otherwise:
\begin{align}
    \nonumber
    H_{trans}(\bm{R}) &= \sum_{1 \leq i < j < k \leq n}H_{trans}^{ijk}(r_{ij}, r_{ik}, r_{jk}) \\
    H_{trans}^{ijk}(r_{ij}, r_{ik}, r_{jk}) &= \delta_{trans}(r_{ik} + r_{ij}r_{jk} - r_{ij}r_{ik} - r_{jk}r_{ik}) \nonumber \\
     &= 
    \begin{cases}
      \delta_{trans}, & [(i \leq j \leq k \leq i) \vee (i \geq j \geq k \geq i)] \\
      0,              & \text{otherwise}
    \end{cases}  \nonumber 
\end{align}
$H_{consist}(\bm{A}, \bm{R})$ is zero if the order encoded in the $\bm{R}$ matrix is consistent with the structure encoded in $\bm{A}$, and $\delta_{consist}$ otherwise.
\begin{align}
    H_{consist}(\bm{A}, \bm{R}) &= \sum_{1 \leq i < j \leq n}H_{consist}^{ij}(a_{ij}, a_{ji}, r_{ij}) \nonumber\\
    H_{consist}^{ij}(a_{ij}, a_{ji}, r_{ij}) &= \delta_{consist}(a_{ji}r_{ij} + a_{ij}(1 - r_{ij})) \nonumber\\
    &=
    \begin{cases}
      \delta_{consist}, & (a_{ji} = r_{ij} = 1) \vee (a_{ij} = 1 \wedge r_{ij} = 0) \\
      0, & \text{otherwise}
    \end{cases}  \nonumber
\end{align}
where $\delta_{trans} \in \mathbb{R^{+}}$ and $\delta_{consist} \in \mathbb{R^{+}}$ are prefixed penalization terms.

\section{\label{sec:qaoa}Quantum Approximate Optimization Algorithm}

Many real optimization problems can be framed as combinatorial problems. The QAOA was proposed by Farhi, Goldstone and Gutmann \cite{farhi2014quantum} for solving combinatorial optimization problems. 

A combinatorial optimization problem is formulated by $n$ bits and $m$ clauses. Each of the clauses affects a subset of bits and is satisfied when this subset is assigned to certain values. Satisfiability asks if a string that satisfies every clause is available. The objective is to maximize the following equation
\begin{equation}
    \label{cost_z}
    C(z) = \sum_{\alpha=1}^{m} C_{\alpha}(z)
\end{equation}
where $z=z_1z_2 \cdots z_n$ is a bit string with $n$ bits and $C_{\alpha}(z)=0$ if clause $\alpha$ is not satisfied by the $z$ string ($C_{\alpha}(z)=1$ otherwise). 

\begin{figure}[H]
\centering
\includegraphics[scale=0.5]{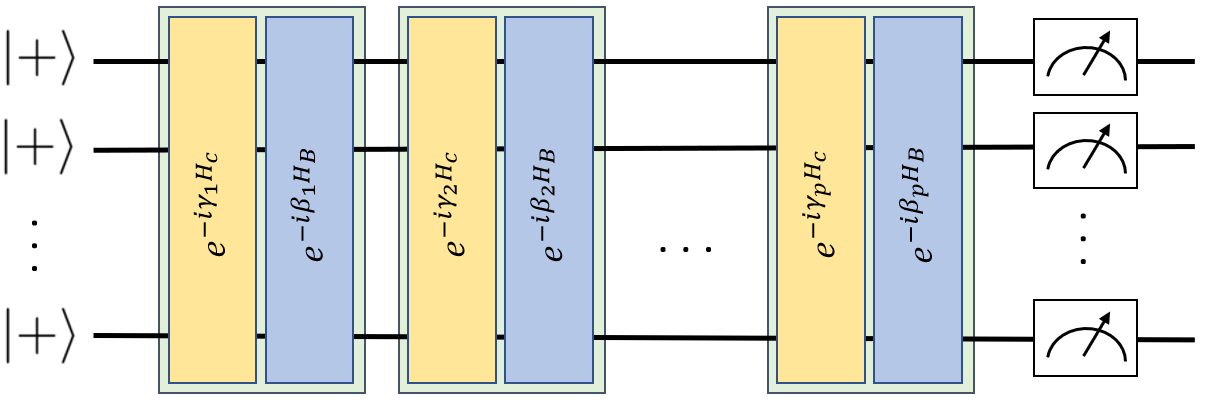}
\caption{A quantum parametric circuit with $p$ layers and $2p$ parameters. The initial state is a superposition of all the possible computational states, and after applying the $p$ layers, a measurement along the $Z$ axis of all the qubits is performed.}
\label{fig:qaoa_circ}
\end{figure}

The QAOA \cite{farhi2014quantum} uses a quantum parametric circuit (variational ansatz) which is built for a specific combinatorial optimization problem and represents a quantum parametric state. In each iteration of the algorithm, the parameters are optimized to find the bit string $z'$ for which $C(z')$ is the maximum of $C$. The quantum circuit consists of $p$ layers, and each layer is formed by two different operators that encode the cost function to be optimized (Fig.~\ref{fig:qaoa_circ}): the cost operator $U(H_C, \gamma)$ parameterized by $\gamma$,
\begin{equation}
    \label{U_c}
    U(H_C, \gamma) = e^{-i\gamma H_C} = \prod_{\alpha=1}^{m} e^{-i\gamma C_{\alpha}}
\end{equation}
and the mixed operator $U(H_B, \beta)$ parameterized by $\beta$,
\begin{equation}
    \label{U_b}
    U(H_B, \beta) = e^{-i\beta H_B} = \prod_{j=1}^{n} e^{-i\beta \sigma_{j}^{x}}
\end{equation}
where $B=\sum_{j=1}^{n}\sigma_{j}^{x}$ and $\sigma_{j}^{x}$ is the mapping of $z_j$ from a binary variable to quantum spin \{+1, -1\}, which is rotated in the $X$-axis. The parameters $\gamma$ and $\beta$ are restricted to lie between $0$ and $2\pi$, as they represent the rotation angles (degrees) over the qubits.

In the literature \cite{behera2020solving, verdon2017quantum}, it has been shown that by increasing the number of layers of a circuit, the algorithm improves its performance and better solutions are obtained. For $p\rightarrow \infty$, the QAOA approximates the adiabatic quantum evolution path \cite{farhi2002quantum}, which is how the algorithm starts from the initial state and converges to a solution. The quantum adiabatic path is approximated in $p$ steps. However, increasing $p$ also increases the depth of the circuit, which entails other disadvantages, such as facing the quantum noise embedded in quantum computers. 
It is then necessary to find the ideal $p$ so as not to drastically increase the depth of the circuit, while still being able to find the final solutions reached by the quantum adiabatic path.

The quantum parametric state is represented as:
\begin{equation}
    \nonumber
    \psi(\bm{\gamma}, \bm{\beta}) = U(H_B, \beta_p) U(H_C, \gamma_p), \dots, U(H_B, \beta_1) U(H_C, \gamma_1) \braket{s}
\end{equation}
where $p\geq 1$, \bm{$\gamma$} = ($\gamma_1, \dots, \gamma_p$), \bm{$\beta$} = ($\beta_1, \dots, \beta_p$), and $\braket{s}$ is the uniform superposition state over all possible computational states. A quantum circuit with $p$ layers and a total of $2p$ parameters ($\gamma_1, \beta_1, \dots, \gamma_p, \beta_p$) to be optimized is shown in Fig.~\ref{fig:qaoa_circ}.

The role of the optimizer is to find the optimal parameters ($\boldsymbol{\gamma}_{opt}$, $\boldsymbol{\beta}_{opt}$) such that the expectation value that encodes the cost function to be optimized,
\begin{equation}
    \nonumber
    f(\bm{\gamma}, \bm{\beta}) = \braket{\psi(\bm{\gamma}, \bm{\beta}) | C | \psi(\bm{\gamma}, \bm{\beta})}
\end{equation}
is minimized. $f(\bm{\gamma}, \bm{\beta})$ encodes the total energy of the system, which should be minimized. Such an expectation can be obtained by performing measurements along the Z-axis of the quantum system and computing the following expression
\begin{equation}
    \label{expectation_value}
    E = \frac{1}{t}\sum_{z\in Z}^{}C(z)N_z
\end{equation}
where $N_z$ is the number of times solution $z$ is measured by executing the circuit $t$ times and $Z$ is the set of possible basis states obtained by the circuit. The total energy of the system, $E$, is expected to be minimized for increasing $p$.

The pseudocode of the QAOA is quite simple once the quantum parametric circuit is built. In each iteration of the algorithm, a new set of (\bm{$\gamma$}, \bm{$\beta$}) parameters is given to the circuit, which is run $t$ times. Then, the optimizer computes the expectation value (Eq.~\eqref{expectation_value}) and proposes a new set of (\bm{$\gamma$}, \bm{$\beta$}) parameters. The loop is repeated until a stopping criterion is met. In each iteration, when the quantum circuit is executed $t$ times, a probability distribution is computed for the solutions. 
The expectation value (Eq.~\eqref{expectation_value}) is also referred to in the machine learning literature as the uncertainty among the solutions, and is expected to be reduced as the algorithm runtime increases. 

Thus, the quantum trial state is prepared, and we then optimize the parameters to bring the trial state as close as possible to the target state. The quality of the QAOA solutions heavily depends on the quality of the parameters (evaluated by Eq.~\eqref{expectation_value}) obtained by the optimizer used during the runtime of the algorithm.

\section{\label{sec:implementation}Method}

This section explains how the variables are deployed in the QAOA approach and how the Hamiltonian is transformed to quantum circuits. All the implemented software is codified by using Qiskit-0.18.1 \cite{Qiskit} and myQLM-1.5.1 \cite{qlm} and is freely available in GitHub\footnote{\url{https://github.com/VicentePerezSoloviev/QAOA_BNSL_IBM}}.

\subsection{QAOA variables}

To make the QAOA able to manage the QUBO variables ($\bm{A}, \bm{R}, \bm{Y}$), it is necessary to arrange them in such a way that they are represented as a vector. Thus, the previous QUBO variables are disposed as a vector of qubits with a size of $v_{size}$ (Eq.~\eqref{number_qubits}).

As explained in Section~\ref{sec:qaoa}, the QAOA is a hybrid approach in which the classic part of the algorithm computes the cost function of the obtained solutions and the expectation value of all the solutions of the corresponding iteration. Our proposal also computes the maximum in-degree ($m = 2$) of the solutions and penalizes those that do not meet the restriction in a classic manner. Upon doing so, $v_{size}$ reduces to
\begin{equation}
    v_{size-QAOA} = n(n-1) + \frac{n(n-1)}{2} = \frac{3n(n-1)}{2},
    \label{number_qubits_opt}
\end{equation}
where $v_{size-QAOA} < v_{size} \; \forall \; n$, because the $\bm{Y}$ variables in the Hamiltonian are not considered.

Thus, the vector $\bm{q}$ needed to solve the BNSL problem for a BN of $n$ nodes by using the QAOA is an array of size $v_{size-QAOA}$,
\begin{equation}
\begin{split}
    \nonumber
    \bm{q} & = (q_1, q_2, \ldots, q_{n*(n-1)} , \ldots, q_{v_{size-QAOA}}) \\
    & = (a_{12}, a_{13}, a_{1n}, \ldots, a_{n(n-1)},
    r_{12}, r_{13}, r_{1n}, r_{23}, \ldots, r_{(n-1) n}),
\end{split}
\end{equation}
where $a_{ij}$ and $r_{ij}$ are defined in Eq.~\eqref{eq_d} and Eq.~\eqref{eq_r}, respectively.

\subsection{QAOA circuit}

In Section~\ref{sec:qaoa}, we have seen that the process of preparing the quantum state during the operation of the QAOA is composed of three elements:
\begin{enumerate}
    \item Preparing an initial state of superposition.
    \item Applying the cost operator $U(H_C, \gamma)$ (Eq.~\eqref{U_c}).
    \item Applying the mixed operator $U(H_B, \beta)$ (Eq.~\eqref{U_b}).
\end{enumerate}

\subsubsection{Initial state}

The initial state used during the QAOA is usually the superposition of all the basis states, which is defined as:
\begin{equation}
    \nonumber
    \ket{\psi_0} = \bigg(\frac{1}{\sqrt{2}}\big(\lvert 0 \rangle + \lvert 1 \rangle\big)\bigg)^{\otimes v_{size-QAOA}},
\end{equation}
where ${\otimes v_{size-QAOA}}$ refers to the number of qubits used in the quantum state (Eq.~\eqref{number_qubits_opt}).

To reach the superposition state of all the possible basis states, we apply Hadamard gates to each qubit ($\pi/2$ degrees in the qubit Y-axis).

\subsubsection{Applying the cost operator $U(H_C, \gamma)$}

As the maximum in-degree verification is implemented in the classic part of the VQA, the Hamiltonian to be implemented is reduced to
\begin{equation}
\label{hamiltonian}
    H(\bm{D}, \bm{R}) = H_{score}(\bm{D}) + H_{trans}(\bm{R}) + H_{consist}(\bm{D}, \bm{R}).
\end{equation}

The Hamiltonian described in Eq.~\eqref{hamiltonian} involves binary variables in $\{0, 1\}$, and the QAOA needs the Hamiltonian to be transformed into a spin Hamiltonian where all the variables are spins in $\{-1, 1\}$. Thus, each binary variable $X_i$ in the QUBO formulation must be transformed as $X_i \rightarrow \frac{1-Z_i}{2}$, where $Z_i$ is the Pauli Z operator that has eigenvalues of $\{-1, +1\}$ and acts on qubit $i$ while ignoring all other qubits:
$$
Z_i =
\begin{pmatrix}
1 & 0 \\
0 & -1 
\end{pmatrix}.
$$

\begin{figure}[h]
\centering
\includegraphics[scale=0.65]{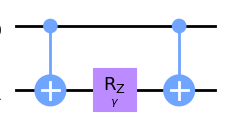}
\caption{A multiplication of two Pauli operators $Z_iZ_j$ is represented in the quantum circuit as a combination of two controlled NOT (CNOT) gates between qubits $i$ and $j$ and a rotation-Z gate in one of them.}
\label{fig:z_iz_j}
\end{figure}

Thus, the QUBO formulation is transformed into a formula in which all the variables involved are $\bm{q}$. The QAOA is a circuit model-based approach, and thus, each Pauli operator $Z_i$ is a quantum gate in the QAOA circuit. Each operator is a rotation-Z gate of qubit $i$, and each multiplication of two Pauli operators $Z_iZ_j$ is a sequence of three gates in qubits $i$ and $j$ (Fig.~\ref{fig:z_iz_j}).

Each $Z_i$ gate has a rotation angle that is parameterized by $\gamma$ and influenced by the structure evaluation scores (Eq.~\eqref{h_score_i}).

\subsubsection{Applying the mixed operator $U(H_B, \beta)$}

The last step of the QAOA circuit is the mixed operator. This operator consists of applying a rotation-X gate in all the qubits of the circuit with parameter $\beta$.


\begin{figure*}[h]
\centering
\includegraphics[width=\textwidth]{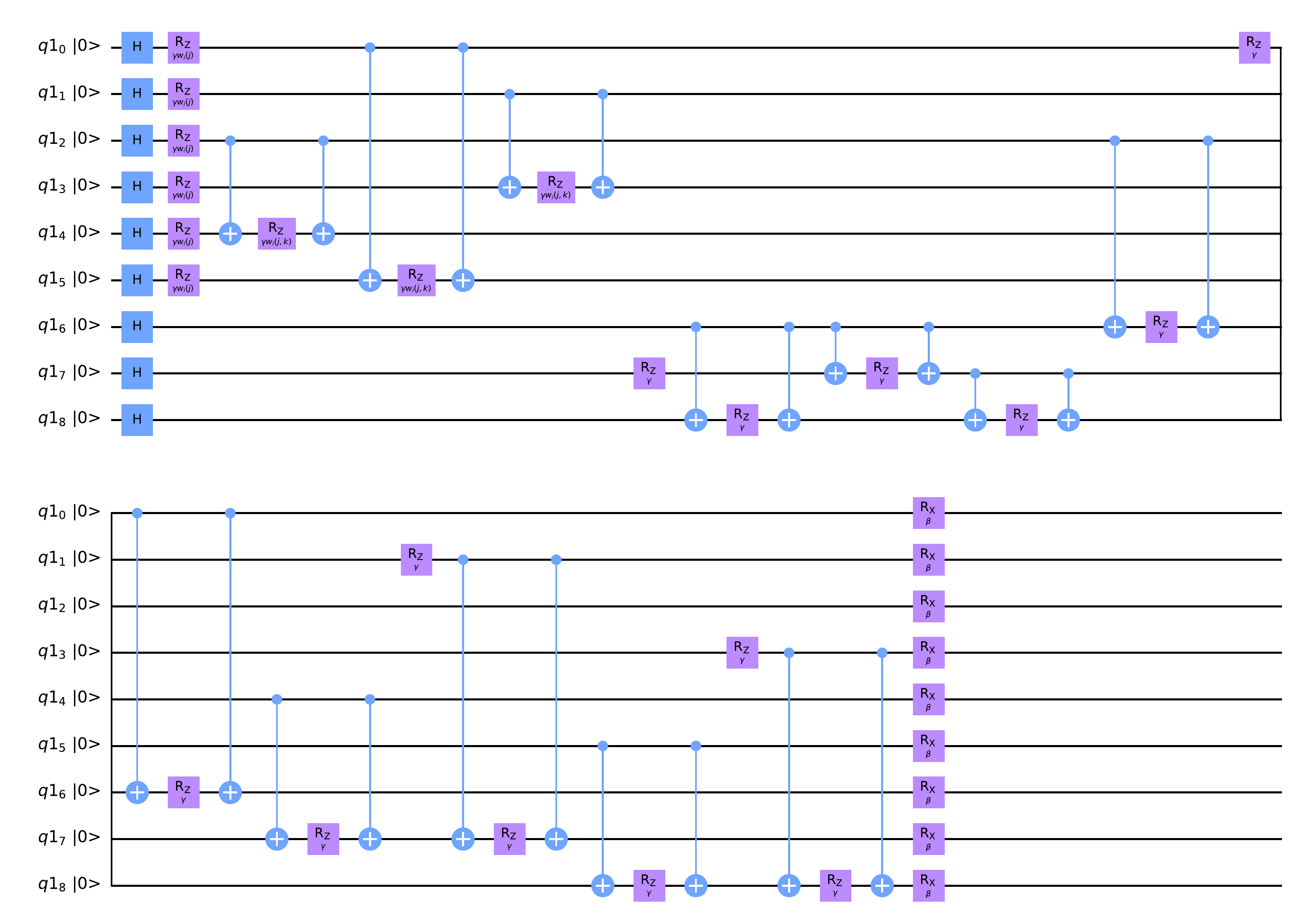}
\caption{A QAOA circuit example for a BNSL problem with 3 nodes and 1 layer. This variational ansatz has $\bm{\gamma}$ and $\bm{\beta}$ parameters as the parameters of the first layer of the circuit.}
\label{fig:qaoa_circ_param}
\end{figure*}

An example of the resultant circuit is shown in Fig.~\ref{fig:qaoa_circ_param}.
For each extra layer, a cost and mixed operators $U(H_C, \bm{\gamma})$ and $U(H_B, \bm{\beta})$ should be added sequentially with their respective parameters to the actual circuit to increase $p$.

\subsection{Conditional value at a risk}
\label{Conditional_Value_at_a_Risk}

As explained in Section~\ref{sec:qaoa}, in each iteration, the expectation value of the measurements along the Z-axis is computed for minimization. This value can be computed classicly by using Eq.~\eqref{expectation_value} after executing the QAOA circuit $t$ times. 

We add a modification to the standard QAOA baseline so that not all the solutions obtained after the measurement process are considered for the expectation value computation. Thus, instead of computing the expectation value, we compute the conditional value at a risk ($CVaR_{\alpha}$), which is a measure that takes only the tail of the distribution of the solutions obtained after measurement into account and is widely used in finance \cite{acerbi2002coherence}. The $CVaR_{\alpha}$ is widely used in different VQAs, such as the QAOA and the variational quantum eigensolver \cite{barkoutsos2020improving}, as it has been proven to lead to faster convergence to better solutions.

Evolutionary algorithms (EAs) \cite{de2016evolutionary} have several characteristics in common with VQAs. A well-known type of EA is the estimation of distribution algorithm \cite{larranaga2001estimation}, which in each iteration sample new solutions from a probability distribution learned from the best old solutions, and then selects the best solutions to update this probability distribution. The top solutions are a percentage of the total set of solutions. This is equivalent to the behaviour of the QAOA considering the $CVaR_{\alpha}$.

A new parameter $\alpha$ is added to the QAOA implementation, and its function is to select the best solutions from the set of solutions measured after executing the QAOA circuit $t$ times. 
Given a cumulative density function $F_{K}$ among all the basis states obtained after measuring the QAOA circuit $t$ times, computing the $CVaR_{\alpha}$ implies computing the expectation value for the $\alpha$-head of $F_{K}$ assuming that samples are sorted in a decreasing order. This selection henceforth referred to as $\alpha K$ with size $\lceil \alpha t \rceil$. Then, the $CVaR_{\alpha}$ is defined as
\begin{equation}
    \label{eq_CVaR_alpha}
   CVaR_{\alpha} =  \frac{1}{\lceil \alpha t \rceil} \sum_{z \in \alpha K} C(z) N_{z_{\alpha K}},
\end{equation}
where $C(z)$ is defined in Eq.~\eqref{cost_z} and $N_{z_{\alpha K}}$ is the number of times that solution $z$ is measured during selection $\alpha K$. The $\alpha$ parameter is defined in the interval $(0, 1]$, such that for $\alpha=1$, the entire set of solutions is considered for the expectation value computation (Eq.~\eqref{expectation_value}), 
and for a decreasing $\alpha$, the number of solutions considered for computing the $CVaR_{\alpha}$ is reduced.

\section{\label{sec:results}Results}

In this section, some results are shown for the BNSL problem after applying the proposed QAOA. Some plots are first given to show how the QAOA performs in terms of cost function minimization (Section~\ref{qaoa_performance}). Then, a perfromance evaluation of the QAOA considering different types of simulated noise is analysed (Section~\ref{noise_resilience}). Finally, a real example of BNSL is shown (Section~\ref{bnsl_data}).

\begin{figure*}[h]
\centering
\includegraphics[scale=0.50]{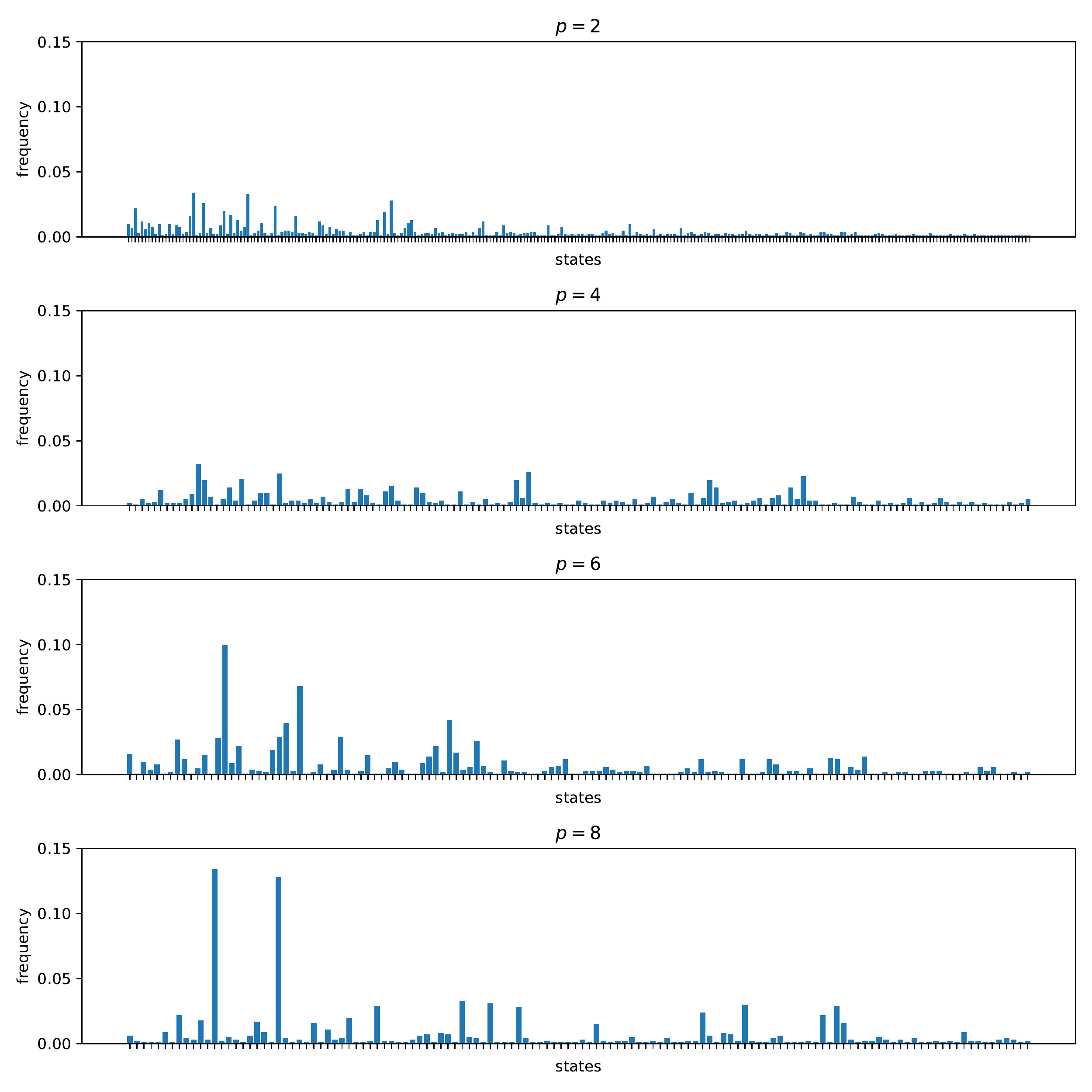}
\caption{Histograms for different numbers of layers $p$ for the same BNSL problem. The $Y$ and $X$ axes represent the frequency, and the solutions, respectively. The names of the solutions have been removed from the $X$-axis for aesthetics, but how they are sorted is the same for each subplot.}
\label{fig:freq_hist_2_4_6_8}
\end{figure*}

Note that the real limitation of this algorithm has been the number of available qubits in the available architectures.
In our experiments, the QAOA looks for the optimal BN structure in a search space containing 543 possible structures for $n=4$ and 29281 structures for $n=5$ (see Eq.~\eqref{possible_structs}). This architecture restriction is imposed due to the number of qubits that we can access at the moment in Qiskit and myQLM. Despite the fact that these problem sizes are far from those examined by the classic BNSL approaches, the number of qubits that companies such as IBM and Google are offering is increasing rapidly, and thus, the use of VQAs is increasingly justified.

The optimizer used in the implementation is the constrained optimization by linear approximation (COBYLA) algorithm \cite{powell1994direct}, which is widely used in the state-of-the-art VQAs \cite{bonet2021performance}.

\subsection{QAOA performance}
\label{qaoa_performance}

\begin{figure*}[h]
\centering
\includegraphics[width=0.90\textwidth]{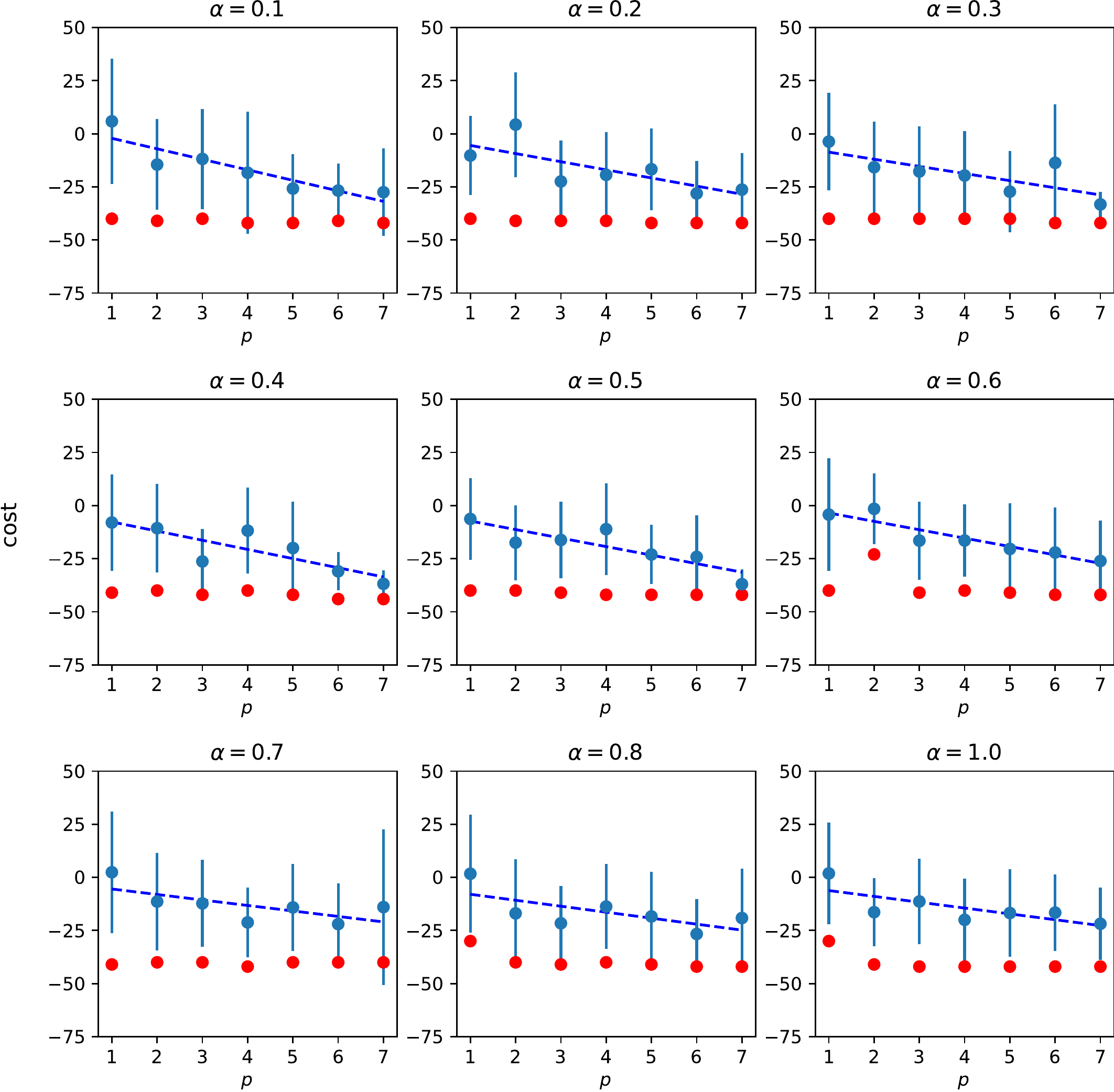}
\caption{Minimization of the $CVaR_{\alpha}$ (Eq.~\eqref{eq_CVaR_alpha}) for different numbers of layers $p$ (X-axis) in the QAOA circuit and different values of the parameter $\alpha$. Blue dots and error bars correspond to the means and standard deviations of the best results found after executing the QAOA 50 times, respectively, and red denotes the minimum costs found in those executions. Dashed trend lines are plotted to guide the human eye.}
\label{fig:plot_n4}
\end{figure*}

The QAOA approach aims at minimizing the uncertainty among the solutions obtained after completing the QAOA circuit measurement process, which is optimized by the minimization of the expectation value; see Eq.~\eqref{expectation_value}. It is expected that by increasing the number of layers $p$ of the circuit, the expectation value among the solutions must decrease. The task of the optimizer is to iteratively search the optimal parameters ($\bm{\gamma}_{opt}$, $\bm{\beta}_{opt}$) of the QAOA circuit to minimize the expectation value. When the optimizer converges to a solution, the parameters ($\bm{\gamma}_{opt}$, $\bm{\beta}_{opt}$) are set to those of the quantum circuit. Fig.~\ref{fig:freq_hist_2_4_6_8} shows an example histogram of the obtained solutions. This experiment is performed with different numbers of layers ($p=2,4,6,8$) to show the differences between the resultant histograms. 

Fig.~\ref{fig:freq_hist_2_4_6_8} shows that increasing the number of layers, clearly minimizes the uncertainty. Note the existence of two clear optima with similar costs for $p=8$; this is not as clear for $p=2$. Moreover, a reduction in the number of solutions that are represented by the $X$-axis for increasing $p$ is clearly visible. The solutions obtained with $p = 2$ have a density close to 0; for $p = 8$, they tend to have a density equal to 0 and thus become insignificant in the corresponding subplot.

As shown before, the QAOA approach is able to reduce the uncertainty among the solutions for the implemented Hamiltonian. However, the proposed approach heavily depends on the random pair of (\bm{$\gamma$}, \bm{$\beta$}) parameters from which the optimizer is initialized. Thus, depending on the initialization, different solutions might be proposed in different executions for the same Hamiltonian problem. 

The minimization of the $CVaR_{\alpha}$ (Eq.~\eqref{eq_CVaR_alpha}) is analysed next by using a random dataset of 4 variables. In Fig.~\ref{fig:plot_n4}, a comparison of the performances achieved by the QAOA for different values of the parameter $\alpha$ and the number of layers $p$ in the QAOA circuit is shown. Note that increasing the number of layers decreases the mean best cost although it increases the depth of the circuit and the computing time.

In Fig.~\ref{fig:plot_n4}, we observe an improvement as $p$ increases and $\alpha$ takes intermediate values. The best solutions are found for intermediate values of $\alpha$ in the range $[0.3, 0.5]$ and $p=7$. 
Note that this improvement with increasing $p$ is not as noticeable for large values of $\alpha$ ($\alpha \rightarrow 1.0$) as it is for the lowest values ($\alpha \rightarrow 0$). 
Despite these results, we claim that it is not necessary to increase the number of layers in the QAOA circuit to find the best results. Fig.~\ref{fig:plot_n4} shows cases in which the same results are obtained, with fewer layers but different values of $\alpha$.

\begin{figure}[h]
\centering
\includegraphics[scale=0.49]{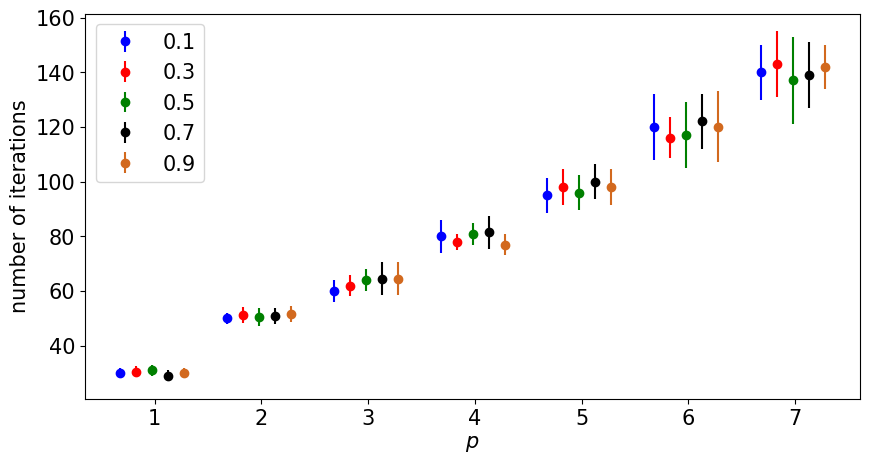}
\caption{Means and standard deviations of the number of iterations until convergence after executing the QAOA 50 times for different values of $p$ (X-axis) and $\alpha$ (colours).}
\label{fig:convergence_alpha}
\end{figure}

In Section~\ref{Conditional_Value_at_a_Risk}, a quantitative comparison of the QAOA approach with estimation of distribution algorithms is provided. This type of algorithm tends to converge to local optima when the percentage of solutions selected to update the probability distribution is too low. However, as shown in Fig.~\ref{fig:plot_n4}, the QAOA does converge to the same solution for any value of $\alpha$. This is analysed in Fig~\ref{fig:convergence_alpha}, where the mean numbers of iterations required until convergence are shown for different executions of the QAOA approach and different values of $p$. Independently of the value of $\alpha$, the number of iterations remains approximately constant for the same value of $p$, whereas it increases with $p$. 

\subsection{Noise resilience}
\label{noise_resilience}

Two main disadvantages of NISQ computers are their limited numbers of qubits and the presence of quantum noise. Thus, there is a need to implement approaches that offer resilience to quantum noise and to optimize the number of qubits used to solve the given problem. It has been shown that VQAs can compensate for quantum errors such as over-/under-rotations \cite{mcclean2016theory}. However, a wide range of studies have analysed the QAOA in different applications to determine the hard limit of its resilience to quantum noise \cite{shaydulin2019evaluating, fontana2021evaluating, urbanek2021mitigating, kandala2019error, sun2021mitigating, vovrosh2021simple}. In other words, we analyse how much noise the QAOA can bear without worsening its optimization behaviour.

\begin{figure*}[h]
\centering
\includegraphics[scale=0.35]{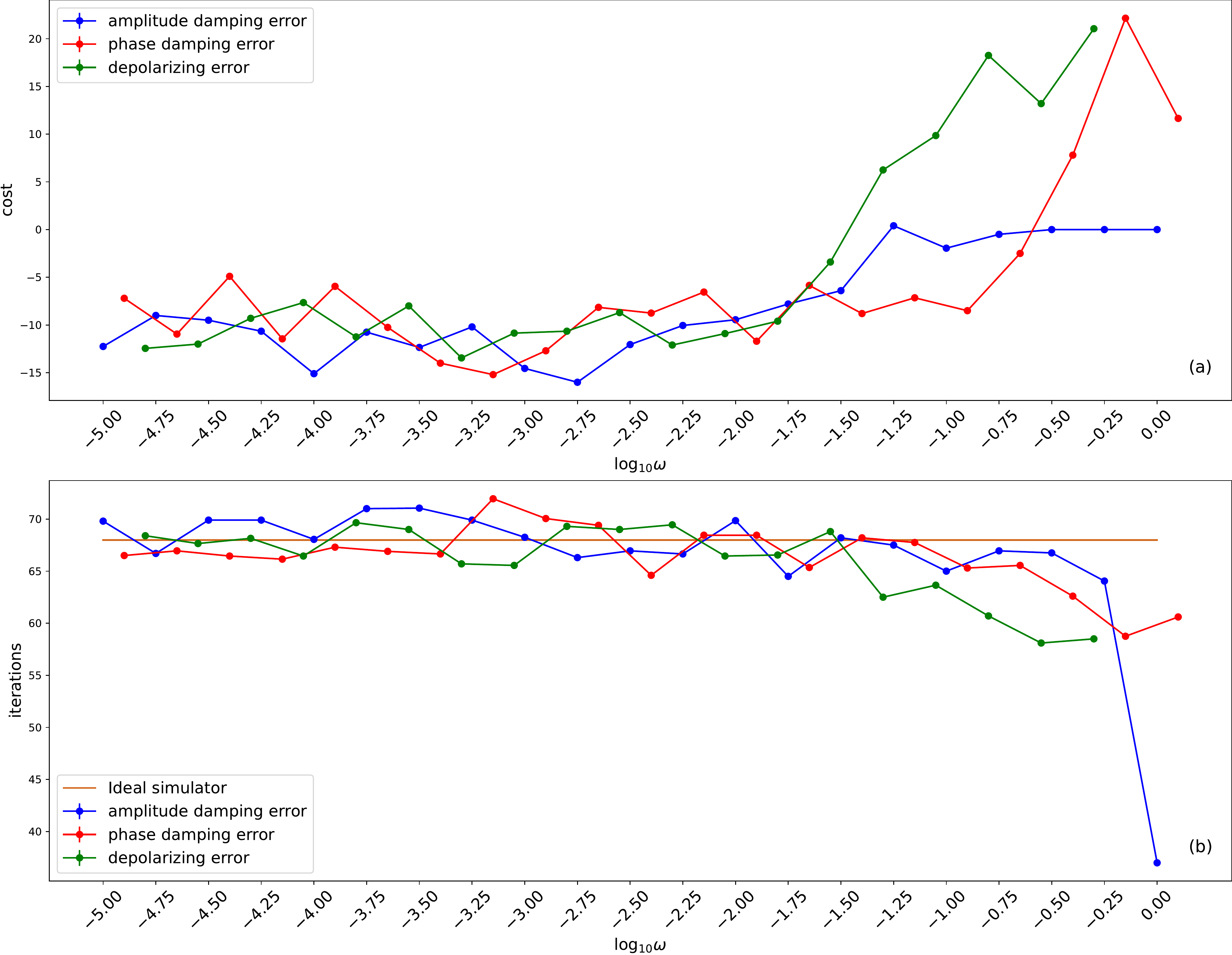}
\caption{Mean best costs (a) and mean numbers of iterations until convergence (b) as a function of the noise amplitude $\omega$. Blue, red and green lines represent the simulated amplitude, phase damping and depolarizing noise models, respectively. For the detailed mean and standard deviation values of these experiments, see the Appendix section. Fifty different executions of the QAOA algorithm were run for the 4 node BNSL problem, and $p=3$.}
\label{fig:noise_cost_conv}
\end{figure*}

To perform this analysis, we construct different noise channels to simulate different types of quantum noises. Two main type of quantum noise can occur: (i) that related to the longitudinal relaxation time (T1) and to the dephasing time (T2) which are metrics for determining the qubit quality; and (ii) that related to the \emph{depolarization noise}. Regarding the former group, the \emph{amplitude damping error} accounts for the loss of energy in a quantum system after T1. After T1, the quantum behaviour is no longer so predictable. On the other hand, the \emph{phase damping error} accounts for the phase loss of a qubit after T2. In this case, the qubit can perform different phase rotations than those it was required to perform. 
Regarding the latter group, the depolarizing error describes the probability that a qubit is depolarized, that is, replaced by a completely mixed state \cite{nielsen2002quantum}.
Other types of quantum noises exist, such as the cross talk error, which is neglected in this study, as the main focus is to analyse the effects of decoherent noise channels (i and ii). Similarly, the readout error is not considered as this type of noise is independent from the ansatz design.

For this study, the noise strength is parameterized by $\omega \in [0,1]$, where $\omega \rightarrow 1$ increases the noise and $\omega = 0$ denotes no noise. The amplitude and damping noises are only applied to the 1-qubit gates, while the depolarization noise is applied to the 2-qubit gates, such as CNOT gates.

Fig.~\ref{fig:noise_cost_conv}(a) shows the resilience of the algorithm to the three types of previously explained noise. The figure shows the mean best costs for different values of $\omega$. Note that the QAOA behaviour remains approximately constant for $\omega \leq 10^{-2}$ with respect to the phase damping error and for $\omega \leq 10^{-2.75}$ with respect to the amplitude damping and depolarizing errors. However, for $\omega \rightarrow 1$, the cost clearly worsens. By analysing the mean best costs obtained from the executions it can be observed that the QAOA exhibits better resilience to the amplitude damping and phase damping errors for larger values of $\omega$, than it does for the depolarizing error. Moreover, the amplitude damping error seems to converge to a stable mean best cost while the other simulated noises worsen with increasing values of $\omega$. 

The numbers of iterations needed by the algorithm to converge are shown in Fig.~\ref{fig:noise_cost_conv}(b). Note that for large values of $\omega$, the number of iterations until convergence decreases. In this case, the QAOA approach converges to local optimal solutions. Thus, the QAOA provides quantitative resilience to quantum errors for approximately $\omega \leq 10^{-2}$. For larger noise values, the algorithm converges to local optimal solutions, as we observe premature convergence to solutions with poor costs. After analysing Fig.~\ref{fig:noise_cost_conv}, we conjecture that our approach has better resilience to the amplitude damping error than to other types of noise. 

\begin{figure*}[h]
\centering
\includegraphics[scale=0.48]{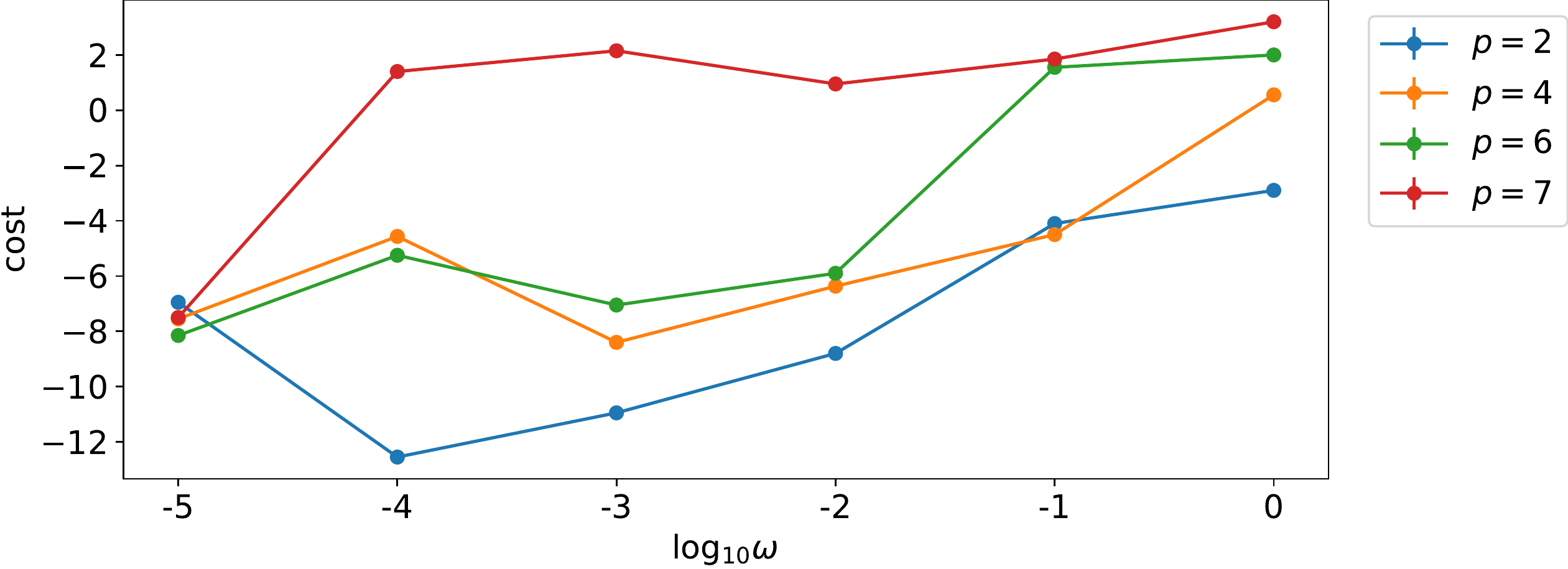}
\caption{Mean best cost obtained for different values of the parameters $\omega$ and $p$ for 50 executions of the QAOA algorithm for the BNSL problem considering the amplitude damping error.}
\label{fig:amplitude_error}
\end{figure*}

In Fig.~\ref{fig:amplitude_error}, a deeper analysis of this noise is shown for different values of $p$. The figure shows the influence of the quantum noise on the results based on the number of layers $p$ of the QAOA circuit. As $p$ increases, the depth of the circuit also increases, and thus, a greater part of the qubit lifespan will be executed outside the coherence times defined by T1 and T2. The results obtained by the QAOA for higher values of $p$ are much worse than those obtained with low values of the parameter, and in most cases a worsening of the mean best cost is observed as $\omega$ increases. From this analysis we conjecture that as the number of layers $p$ increases, the QAOA becomes less resilient to the amplitude damping error. 

\subsection{BNSL from real-world data}
\label{bnsl_data}

\begin{figure*}[h]
\centering
\includegraphics[width=0.30\textwidth]{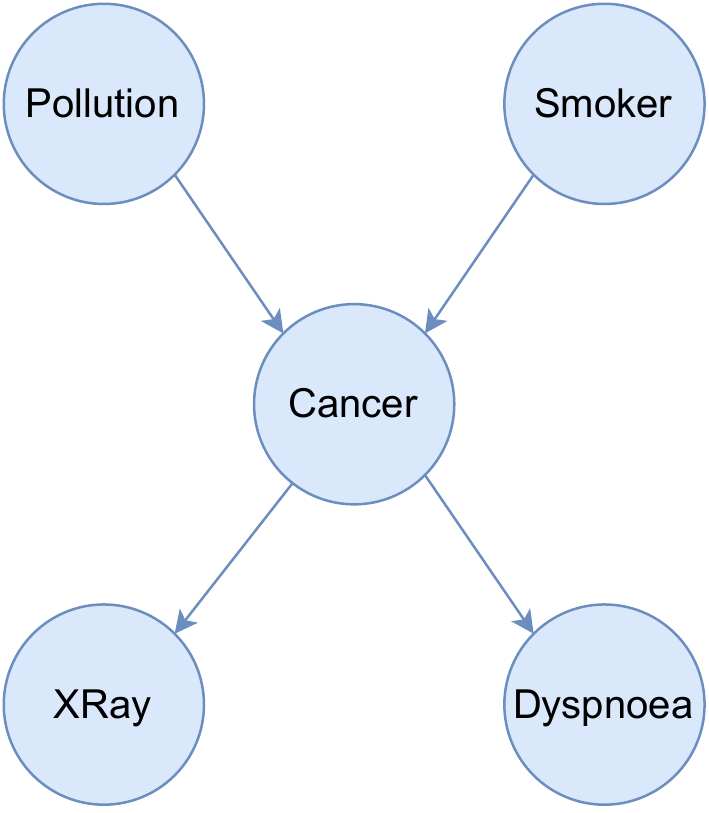}
\caption{Original \texttt{Cancer} BN structure composed of 5 nodes that represent 5 discrete variables and 4 arcs.}
\label{fig:cancer_bn}
\end{figure*}

In this section, the QAOA approach is applied to a real BNSL problem by using the \texttt{Cancer}\footnote{\url{https://www.bnlearn.com/bnrepository/discrete-small.html\#cancer}} benchmark. The \texttt{Cancer} BN (Fig.~\ref{fig:cancer_bn}) has 5 discrete variables, from which we sampled 3 different datasets using probabilistic logic sampling \cite{henrion1988propagating} with 500, 1000 and 10000 instances. The structures provided by the QAOA are compared to the original BN structure through the structural Hamming distance (\textit{SHD}) metric, where \textit{SHD}${=}0$ means that the QAOA approach fully recovers all the arcs of the original BN. We consider two different structure evaluation scores: the BIC and BDeu scores. This experiment is limited to 5 nodes due to the limit of qubits we can access with Qiskit and myQLM, for which the search space is composed of 29281 possible BN structures.

\begin{table}[ht]
\caption{Comparison of the QAOA approach with three different classic approaches and simulated quantum annealing. The experiment is executed 10 times for each of the non-deterministic (SQA and QAOA) approaches (the best results are shown), and three different dataset sizes are simulated with 500, 1000 and 10000 instances. BIC and BDeu scores are used for the local BN structure evaluation. The \textit{SHD} metric is shown for each experiment.}
\vspace*{5mm}
\centering

\begin{tabular}{|c|c|c|c|c|c|c|c|}

\hline
&   & \multicolumn{2}{c|}{500} & \multicolumn{2}{c|}{1000} & \multicolumn{2}{c|}{10000}\\
\hline
{} & $\alpha$ & BIC & BDeu & BIC & BDeu & BIC & BDeu \\
\hline
HC        &     - &       4 &        4 &        4 &         4 &         \cellcolor{blue!25}0 &          \cellcolor{blue!25}0 \\
Tabu      &     - &       4 &        4 &        4 &         5 &         \cellcolor{blue!25}0 &          \cellcolor{blue!25}0 \\
MMHC      &     - &       4 &        4 &        4 &         4 &         \cellcolor{blue!25}0 &          \cellcolor{blue!25}0 \\
SQA       &     - &       5 &        5 &        4 &         5 &         3 &          2 \\
QAOA      &   0.9 &       \cellcolor{blue!25}0 &        \cellcolor{blue!25}0 &        1 &         \cellcolor{blue!25}0 &         \cellcolor{blue!25}0 &          1 \\
QAOA      &   0.7 &       1 &        1 &        1 &         \cellcolor{blue!25}0 &         \cellcolor{blue!25}0 &          \cellcolor{blue!25}0 \\
QAOA      &   0.5 &       1 &        1 &        \cellcolor{blue!25}0 &         1 &         1 &          1 \\
QAOA      &   0.3 &       \cellcolor{blue!25}0 &        \cellcolor{blue!25}0 &        1 &         \cellcolor{blue!25}0 &         \cellcolor{blue!25}0 &          1 \\
\hline
\end{tabular}
\label{tab:cancer_comparison}
\end{table}

We compare the results obtained by the QAOA approach with those of two score-based algorithms: the hill climbing (HC) \cite{Hill_climbing} and tabu search \cite{tabu} algorithms; with that of a hybrid algorithm: max-min hill climbing (MMHC) \cite{max-min-hill-climbing}; and with that of the simulated quantum annealing (SQA), which is a version of quantum annealing executed in the quantum learning machine \cite{qlm}. The QAOA results are shown in Table~\ref{tab:cancer_comparison} for different values of $\alpha$, where the number of layers $p$ is optimized for each case. The parameters of the algorithms are optimized, and the best results are shown in Table~\ref{tab:cancer_comparison}.

The QAOA approach obtains better results than those of classic approaches regardless of the $\alpha$ parameter for a low number of instances. As the number of instances increases, the $\alpha$ parameter seems to have a larger influence on the obtained results. The QAOA improves upon the results of the SQA in all the experiments.

\section{Conclusions}
\label{conclusions}

In this work the BNSL problem was approached by using the QAOA quantum variational algorithm. The problem was transformed into a Hamiltonian energy function, and then translated into a QAOA parametric circuit to be optimized into a classic loop. 

A remarkable uncertainty reduction across all the possible solutions was shown for increasing number of QAOA circuit layers. We introduced the concept of the $CVaR_{\alpha}$ to reduce the number of solutions selected for computing the expectation value of each iteration. Considering this new parameter, we could observe that it was not necessary to increase the number of circuit layers to obtain the best results since the QAOA was able to converge to similar solutions by tuning the $\alpha$ parameter.

The NISQ-era quantum computers are characterized by the quantum noise embedded in these systems. We analysed the performance of our approach while simulating different types of quantum noise, and our results show that the QAOA is resilient to quantum noise over a range of noise amplitudes. More specifically, our approach offers better performance when considering the amplitude damping error, which was more deeply analyzed.

Our approach was also applied to the \texttt{Cancer} benchmark and a comparison with other optimizers was shown with different structure evaluation scores and dataset sizes. The QAOA found the global optimum for every size-score combination and seemed to outperform classic and quantum approaches on any dataset.

Considering the results obtained in this work, we believe that the use of VQAs to solve the BNSL problem is justified. As future lines of research, we suggest considering the warm starting scenario \cite{egger2021warm} and a more in-depth analysis of other quantum noises with different quantum computers. 
Even though the sizes of the BNs that the QAOA has faced are small, we believe that this approach will be worthwhile for further research as the number of qubits that can be freely accessed increases.

\appendix
\section{Appendices}
\label{app_tab}

Table~\ref{tab:cost_noise} shows the detailed mean and standard deviation of the experimental results represented in Fig.~\ref{fig:noise_cost_conv}.

\input{table_noise}

\section*{Acknowledgements}
The authors would like to thank A. Gomez from Centro de Supercomputacion de Galicia (CESGA) for helpful discussions. 
We would like to thank Cestro Singular de Investigación en Tecnoloxías Intelixentes (CITIUS) and CESGA for access to the computers where the experiments were carried out.
We acknowledge the access to advanced services provided by the IBM Quantum Researchers Program.
This work has been partially supported by the Spanish Ministry of Science and
Innovation through the PID2019-109247GB-I00 and RTC2019-006871-7 projects, and by the
BBVA Foundation (2019 Call) through the ``Score-based nonstationary temporal Bayesian
networks. Applications in climate and neuroscience" (BAYES-CLIMA-NEURO) project.
Vicente P. Soloviev has been supported by the FPI PRE2020-094828 PhD grant from the Spanish Ministry of Science and Innovation.

\bibliographystyle{plainnat}
\bibliography{mybibfile}

\end{document}

%% file: table_noise.tex
\begin{table}[H]
    \centering
    \addtolength{\leftskip} {-2cm}
    \addtolength{\rightskip}{-2cm}
    \begin{tabular}{|c|c|c|c|c|c|c|c|c|c|c|c|c|c|}
    \hline
    & \multicolumn{6}{c|}{cost} & \multicolumn{6}{c|}{convergence}  \\
    \hline
    & \multicolumn{2}{c|}{AD} & \multicolumn{2}{c|}{PD} & \multicolumn{2}{c|}{DE} & \multicolumn{2}{c|}{AD} & \multicolumn{2}{c|}{PD} & \multicolumn{2}{c|}{DE} \\
    \hline
    $log_{10}\omega$ &  $\mu$ &  $\sigma$ &  $\mu$ &  $\sigma$ &  $\mu$ &  $\sigma$ & $\mu$ &  $\sigma$ &  $\mu$ &  $\sigma$ &  $\mu$ &  $\sigma$ \\
    \hline
    -5.00 &        -12.25 &        10.80 &         -7.20 &        15.61 &        -12.45 &         8.96 &         69.80 &         4.70 &         66.50 &         6.73 &         68.40 &         6.64 \\
    -4.75 &         -9.00 &        10.53 &        -10.95 &        10.80 &        -12.00 &         7.94 &         66.70 &         6.34 &         66.95 &         5.13 &         67.65 &         6.04 \\
    -4.50 &         -9.50 &        10.69 &         -4.90 &        11.49 &         -9.30 &         9.65 &         69.90 &         7.79 &         66.45 &         8.11 &         68.15 &         7.80 \\
    -4.25 &        -10.65 &         8.97 &        -11.45 &        10.95 &         -7.65 &        11.34 &         69.90 &         7.83 &         66.15 &         5.95 &         66.45 &         5.36 \\
    -4.00 &        -15.10 &         7.75 &         -5.95 &         9.58 &        -11.25 &        11.50 &         68.05 &         5.48 &         67.30 &         5.53 &         69.65 &         4.36 \\
    -3.75 &        -10.75 &        10.97 &        -10.25 &        11.54 &         -8.00 &        12.53 &         71.00 &         9.00 &         66.90 &         5.18 &         69.00 &         5.52 \\
    -3.50 &        -12.35 &         9.17 &        -14.00 &         8.47 &        -13.45 &         7.47 &         71.05 &         8.01 &         66.65 &         5.65 &         65.70 &         6.06 \\
    -3.25 &        -10.20 &        13.78 &        -15.20 &         8.79 &        -10.85 &        10.98 &         69.90 &         8.47 &         71.95 &         7.16 &         65.55 &         6.07 \\
    -3.00 &        -14.55 &         7.55 &        -12.70 &         8.86 &        -10.65 &         8.64 &         68.25 &         7.35 &         70.05 &         9.41 &         69.30 &         5.92 \\
    -2.75 &        -16.00 &         8.07 &         -8.15 &        13.69 &         -8.70 &        11.72 &         66.30 &         5.20 &         69.40 &         7.61 &         69.00 &         6.28 \\
    -2.50 &        -12.05 &         8.84 &         -8.75 &        13.18 &        -12.10 &        10.34 &         66.95 &         5.52 &         64.60 &         6.94 &         69.45 &         7.94 \\
    -2.25 &        -10.05 &         9.40 &         -6.55 &        12.15 &        -10.90 &         8.23 &         66.65 &         4.69 &         68.45 &         6.63 &         66.45 &         6.37 \\
    -2.00 &         -9.45 &        11.91 &        -11.70 &         9.59 &         -9.60 &        12.33 &         69.85 &         6.18 &         68.45 &         6.10 &         66.55 &         5.07 \\
    -1.75 &         -7.80 &        12.76 &         -5.85 &        12.78 &         -3.40 &         9.33 &         64.50 &         5.84 &         65.35 &         5.29 &         68.80 &         7.70 \\
    -1.50 &         -6.40 &         7.16 &         -8.80 &        11.62 &          6.25 &        15.34 &         68.20 &         5.52 &         68.20 &         6.96 &         62.50 &         4.88 \\
    -1.25 &          0.40 &         9.30 &         -7.15 &        10.16 &          9.85 &        16.68 &         67.50 &         6.06 &         67.75 &         6.22 &         63.65 &         5.58 \\
    -1.00 &         -1.95 &         5.67 &         -8.50 &        10.04 &         18.25 &        13.26 &         65.00 &         6.05 &         65.30 &         6.39 &         60.70 &         4.66 \\
    -0.75 &         -0.50 &         2.24 &         -2.50 &         7.25 &         13.20 &        12.89 &         66.95 &         6.48 &         65.55 &         5.86 &         58.10 &         4.59 \\
    -0.50 &          0.00 &         0.00 &          7.80 &        12.88 &         21.05 &        11.81 &         66.75 &         5.76 &         62.60 &         3.76 &         58.50 &         4.57 \\
    -0.25 &          0.00 &         0.00 &         22.15 &        14.53 &         20.01 &        10.30 &         64.05 &         6.73 &         58.75 &         4.02 &         59.10 &         4.59 \\
     0.00 &          0.00 &         0.00 &         11.65 &        13.35 &         21.50 &        11.20 &         37.00 &         0.00 &         60.60 &         5.86 &         58.60 &         4.80 \\
     \hline
    \end{tabular}
    \caption{Mean best costs ($\mu$) and standard deviations ($\sigma$) found for different values of the $\omega$ parameter over 50 executions of the QAOA approach for the BNSL problem, and the mean numbers of iterations ($\mu$) and standard deviations ($\sigma$) until convergence for 50 executions. AD, PD and DE represent amplitude damping, phase damping, and depolarizing simulated errors, respectively.}
    \label{tab:cost_noise}
\end{table}

%% file: main.bbl
\begin{thebibliography}{47}
\providecommand{\natexlab}[1]{#1}
\providecommand{\url}[1]{\texttt{#1}}
\expandafter\ifx\csname urlstyle\endcsname\relax
  \providecommand{\doi}[1]{doi: #1}\else
  \providecommand{\doi}{doi: \begingroup \urlstyle{rm}\Url}\fi

\bibitem[Acerbi and Tasche(2002)]{acerbi2002coherence}
Carlo Acerbi and Dirk Tasche.
\newblock On the coherence of expected shortfall.
\newblock \emph{\textit{Journal of Banking and Finance}}, 26\penalty0
  (7):\penalty0 1487--1503, 2002.

\bibitem[Aleksandrowicz et~al.(2021)Aleksandrowicz, Alexander, Barkoutsos,
  Bello, Ben-Haim, Bucher, Cabrera-Hern{\'a}ndez, Carballo-Franquis, Chen,
  Chen, et~al.]{Qiskit}
Gadi Aleksandrowicz, Thomas Alexander, Panagiotis Barkoutsos, Luciano Bello,
  Yael Ben-Haim, David Bucher, Francisco~Jose Cabrera-Hern{\'a}ndez, Jorge
  Carballo-Franquis, Adrian Chen, Chun-Fu Chen, et~al.
\newblock {Qiskit: An Open-source Framework for Quantum Computing}, 2021.

\bibitem[Aouay et~al.(2013)Aouay, Jamoussi, and Ayed]{aouay2013particle}
Saoussen Aouay, Salma Jamoussi, and Yassine~Ben Ayed.
\newblock {Particle swarm optimization based method for Bayesian network
  structure learning}.
\newblock In \emph{2013 5th International Conference on Modeling, Simulation
  and Applied Optimization}, pages 1--6. IEEE, 2013.

\bibitem[ATOS(2021)]{qlm}
ATOS.
\newblock {Quantum Learning Machine}.
\newblock \url{https://atos.net/en/solutions/quantum-learning-machine}, 2021.
\newblock [Online; accessed 26-January-2022].

\bibitem[Barkoutsos et~al.(2020)Barkoutsos, Nannicini, Robert, Tavernelli, and
  Woerner]{barkoutsos2020improving}
Panagiotis~Kl Barkoutsos, Giacomo Nannicini, Anton Robert, Ivano Tavernelli,
  and Stefan Woerner.
\newblock {Improving variational quantum optimization using CVaR}.
\newblock \emph{\textit{Quantum}}, 4:\penalty0 256, 2020.

\bibitem[Bielza and Larrañaga(2014)]{bielza2014bayesian}
Concha Bielza and Pedro Larrañaga.
\newblock {Bayesian networks in neuroscience: A survey}.
\newblock \emph{\textit{Frontiers in Computational Neuroscience}}, 8:\penalty0
  131, 2014.

\bibitem[Blanco et~al.(2003)Blanco, Inza, and Larrañaga]{blanco2003learning}
Rosa Blanco, Inaki Inza, and Pedro Larrañaga.
\newblock {Learning Bayesian networks in the space of structures by estimation
  of distribution algorithms}.
\newblock \emph{\textit{{International Journal of Intelligent Systems}}},
  18\penalty0 (2):\penalty0 205--220, 2003.

\bibitem[Bonet-Monroig et~al.(2021)Bonet-Monroig, Wang, Vermetten, Senjean,
  Moussa, B{\"a}ck, Dunjko, and O'Brien]{bonet2021performance}
Xavier Bonet-Monroig, Hao Wang, Diederick Vermetten, Bruno Senjean, Charles
  Moussa, Thomas B{\"a}ck, Vedran Dunjko, and Thomas~E O'Brien.
\newblock Performance comparison of optimization methods on variational quantum
  algorithms.
\newblock \emph{\textit{arXiv preprint arXiv:2111.13454}}, 2021.

\bibitem[Chickering(1996)]{chickering1996learning}
David~Maxwell Chickering.
\newblock Learning bayesian networks is np-complete.
\newblock In \emph{Learning from data}, pages 121--130. Springer, 1996.

\bibitem[Choi and Kim(2019)]{choi2019tutorial}
Jaeho Choi and Joongheon Kim.
\newblock {A tutorial on quantum approximate optimization algorithm (QAOA):
  Fundamentals and Applications}.
\newblock In \emph{2019 International Conference on Information and
  Communication Technology Convergence}, pages 138--142. IEEE, 2019.

\bibitem[Cooper and Herskovits(1992)]{cooper1992bayesian}
Gregory~F Cooper and Edward Herskovits.
\newblock {A Bayesian method for the induction of probabilistic networks from
  data}.
\newblock \emph{\textit{Machine learning}}, 9\penalty0 (4):\penalty0 309--347,
  1992.

\bibitem[De~Jong(2016)]{de2016evolutionary}
Kenneth De~Jong.
\newblock {Evolutionary computation: A unified approach}.
\newblock In \emph{Proceedings of the 2016 Genetic and Evolutionary Computation
  Conference Companion}, pages 185--199. {The MIT Press}, 2016.

\bibitem[Egger et~al.(2021)Egger, Mare{\v{c}}ek, and Woerner]{egger2021warm}
Daniel~J Egger, Jakub Mare{\v{c}}ek, and Stefan Woerner.
\newblock Warm-starting quantum optimization.
\newblock \emph{\textit{Quantum}}, 5:\penalty0 479, 2021.

\bibitem[Farhi et~al.(2002)Farhi, Goldstone, and Gutmann]{farhi2002quantum}
Edward Farhi, Jeffrey Goldstone, and Sam Gutmann.
\newblock Quantum adiabatic evolution algorithms with different paths.
\newblock \emph{\textit{arXiv preprint quant-ph/0208135}}, 2002.

\bibitem[Farhi et~al.(2014)Farhi, Goldstone, and Gutmann]{farhi2014quantum}
Edward Farhi, Jeffrey Goldstone, and Sam Gutmann.
\newblock A quantum approximate optimization algorithm.
\newblock \emph{\textit{arXiv preprint arXiv:1411.4028}}, 2014.

\bibitem[Fontana et~al.(2021)Fontana, Fitzpatrick, Ramo, Duncan, and
  Rungger]{fontana2021evaluating}
Enrico Fontana, Nathan Fitzpatrick, David~Mu{\~n}oz Ramo, Ross Duncan, and Ivan
  Rungger.
\newblock Evaluating the noise resilience of variational quantum algorithms.
\newblock \emph{\textit{{Physical Review A}}}, 104\penalty0 (2):\penalty0
  022403, 2021.

\bibitem[Gámez et~al.(2011)Gámez, Mateo, and Puerta]{Hill_climbing}
José~A. Gámez, Juan Mateo, and Jose~M. Puerta.
\newblock {Learning Bayesian networks by hill climbing: Efficient methods based
  on progressive restriction of the neighborhood}.
\newblock \emph{\textit{Data Mining and Knowledge Discovery}}, 22:\penalty0
  106--148, 2011.

\bibitem[Hauke et~al.(2020)Hauke, Katzgraber, Lechner, Nishimori, and
  Oliver]{hauke2020perspectives}
Philipp Hauke, Helmut~G Katzgraber, Wolfgang Lechner, Hidetoshi Nishimori, and
  William~D Oliver.
\newblock {Perspectives of quantum annealing: Methods and implementations}.
\newblock \emph{Reports on Progress in Physics}, 83\penalty0 (5):\penalty0
  054401, 2020.

\bibitem[Heckerman et~al.(1995)Heckerman, Geiger, and Chickering]{BDe}
David Heckerman, Dan Geiger, and David~M Chickering.
\newblock {Learning Bayesian networks: The combination of knowledge and
  statistical data}.
\newblock \emph{\textit{{Machine Learning}}}, 20\penalty0 (3):\penalty0
  197--243, 1995.

\bibitem[Henrion(1988)]{henrion1988propagating}
Max Henrion.
\newblock {Propagating uncertainty in Bayesian networks by probabilistic logic
  sampling}.
\newblock In \emph{{Machine Intelligence and Pattern Recognition}}, volume~5,
  pages 149--163. Elsevier, 1988.

\bibitem[Ji et~al.(2011)Ji, Zhang, Hu, and Liu]{tabu}
Jun-Zhong Ji, Hong-Xun Zhang, Ren-Bing Hu, and Chun-Nian Liu.
\newblock {A tabu-search based Bayesian network structure learning algorithm}.
\newblock \emph{\textit{Journal of Beijing University of Technology}},
  37:\penalty0 1274--1280, 2011.

\bibitem[Kandala et~al.(2019)Kandala, Temme, C{\'o}rcoles, Mezzacapo, Chow, and
  Gambetta]{kandala2019error}
Abhinav Kandala, Kristan Temme, Antonio~D C{\'o}rcoles, Antonio Mezzacapo,
  Jerry~M Chow, and Jay~M Gambetta.
\newblock Error mitigation extends the computational reach of a noisy quantum
  processor.
\newblock \emph{\textit{Nature}}, 567\penalty0 (7749):\penalty0 491--495, 2019.

\bibitem[Koller and Friedman(2009)]{koller2009probabilistic}
Daphne Koller and Nir Friedman.
\newblock \emph{\textit{{Probabilistic Graphical Models: Principles and
  Techniques}}}.
\newblock The MIT Press, 2009.

\bibitem[Larra{\~n}aga and Lozano(2001)]{larranaga2001estimation}
Pedro Larra{\~n}aga and Jose~A Lozano.
\newblock \emph{{Estimation of Distribution Algorithms: A New Tool for
  Evolutionary Computation}}, volume~2.
\newblock Springer, 2001.

\bibitem[Larrañaga et~al.(1996)Larrañaga, Poza, Yurramendi, Murga, and
  Kuijpers]{larranaga1996structure}
Pedro Larrañaga, Mikel Poza, Yosu Yurramendi, Roberto~H. Murga, and Cindy
  M.~H. Kuijpers.
\newblock {Structure learning of Bayesian networks by genetic algorithms: A
  performance analysis of control parameters}.
\newblock \emph{\textit{{IEEE Transactions on Pattern Analysis and Machine
  Intelligence}}}, 18\penalty0 (9):\penalty0 912--926, 1996.

\bibitem[Lee and Kim(2019)]{lee2019parallel}
Sangmin Lee and Seoung~Bum Kim.
\newblock {Parallel simulated annealing with a greedy algorithm for Bayesian
  network structure learning}.
\newblock \emph{IEEE Transactions on Knowledge and Data Engineering},
  32\penalty0 (6):\penalty0 1157--1166, 2019.

\bibitem[McClean et~al.(2016)McClean, Romero, Babbush, and
  Aspuru-Guzik]{mcclean2016theory}
Jarrod~R McClean, Jonathan Romero, Ryan Babbush, and Al{\'a}n Aspuru-Guzik.
\newblock The theory of variational hybrid quantum-classical algorithms.
\newblock \emph{\textit{New Journal of Physics}}, 18\penalty0 (2):\penalty0
  023023, 2016.

\bibitem[Murphy(2012)]{murphy2012machine}
Kevin~P Murphy.
\newblock \emph{\textit{{Machine Learning: A Probabilistic Perspective}}}.
\newblock The MIT press, 2012.

\bibitem[Nielsen and Chuang(2002)]{nielsen2002quantum}
Michael~A Nielsen and Isaac Chuang.
\newblock \emph{\textit{Quantum Computation and Quantum Information}}.
\newblock American Association of Physics Teachers, 2002.

\bibitem[O’Gorman et~al.(2015)O’Gorman, Babbush, Perdomo-Ortiz,
  Aspuru-Guzik, and Smelyanskiy]{o2015bayesian}
Bryan O’Gorman, Ryan Babbush, Alejandro Perdomo-Ortiz, Al{\'a}n Aspuru-Guzik,
  and Vadim Smelyanskiy.
\newblock Bayesian network structure learning using quantum annealing.
\newblock \emph{\textit{The European Physical Journal Special Topics}},
  224\penalty0 (1):\penalty0 163--188, 2015.

\bibitem[Peruzzo et~al.(2014)Peruzzo, McClean, Shadbolt, Yung, Zhou, Love,
  Aspuru-Guzik, and O’Brien]{peruzzo2014variational}
Alberto Peruzzo, Jarrod McClean, Peter Shadbolt, Man-Hong Yung, Xiao-Qi Zhou,
  Peter~J Love, Al{\'a}n Aspuru-Guzik, and Jeremy~L O’Brien.
\newblock A variational eigenvalue solver on a photonic quantum processor.
\newblock \emph{\textit{{Nature Communications}}}, 5\penalty0 (1):\penalty0
  1--7, 2014.

\bibitem[Powell(1994)]{powell1994direct}
Michael~JD Powell.
\newblock A direct search optimization method that models the objective and
  constraint functions by linear interpolation.
\newblock In \emph{{Advances in Optimization and Numerical Analysis}}, pages
  51--67. Springer, 1994.

\bibitem[Puerto-Santana et~al.(2021)Puerto-Santana, Larrañaga, and
  Bielza]{9387117}
Carlos Puerto-Santana, Pedro Larrañaga, and Concha Bielza.
\newblock {Autoregressive asymmetric linear Gaussian hidden Markov models}.
\newblock \emph{\textit{{IEEE Transactions on Pattern Analysis and Machine
  Intelligence}}}, 2021.

\bibitem[Quesada et~al.(2021)Quesada, Bielza, and
  Larra{\~n}aga]{quesada2021structure}
David Quesada, Concha Bielza, and Pedro Larra{\~n}aga.
\newblock {Structure learning of high-order dynamic Bayesian networks via
  particle swarm optimization with order invariant encoding}.
\newblock In \emph{International Conference on Hybrid Artificial Intelligence
  Systems}, pages 158--171. Springer, 2021.

\bibitem[Robinson(1977)]{robinson1977counting}
Robert~W Robinson.
\newblock Counting unlabeled acyclic digraphs.
\newblock In \emph{\textit{Combinatorial Mathematics V}}, pages 28--43.
  Springer, 1977.

\bibitem[Scanagatta et~al.(2019)Scanagatta, Salmer{\'o}n, and
  Stella]{scanagatta2019survey}
Mauro Scanagatta, Antonio Salmer{\'o}n, and Fabio Stella.
\newblock {A survey on Bayesian network structure learning from data}.
\newblock \emph{\textit{Progress in Artificial Intelligence}}, 8\penalty0
  (4):\penalty0 425--439, 2019.

\bibitem[Schuld and Petruccione(2018)]{schuld2018supervised}
Maria Schuld and Francesco Petruccione.
\newblock \emph{\textit{Supervised Learning with Quantum Computers}}.
\newblock Springer, 2018.

\bibitem[Schwarz(1978)]{schwarz1978estimating}
Gideon Schwarz.
\newblock Estimating the dimension of a model.
\newblock \emph{\textit{The Annals of Statistics}}, pages 461--464, 1978.

\bibitem[Shaydulin and Alexeev(2019)]{shaydulin2019evaluating}
Ruslan Shaydulin and Yuri Alexeev.
\newblock Evaluating quantum approximate optimization algorithm: A case study.
\newblock In \emph{{2019 Tenth International Green and Sustainable Computing
  Conference}}, pages 1--6. IEEE, 2019.

\bibitem[Shikuri(2020)]{shikuri2020efficient}
Yuta Shikuri.
\newblock {Efficient conversion of Bayesian network learning into quadratic
  unconstrained binary optimization}.
\newblock \emph{\textit{arXiv preprint arXiv:2006.06926}}, 2020.

\bibitem[Streif and Leib(2019)]{streif2019comparison}
Michael Streif and Martin Leib.
\newblock {Comparison of QAOA with quantum and simulated annealing}.
\newblock \emph{\textit{arXiv preprint arXiv:1901.01903}}, 2019.

\bibitem[Sun et~al.(2021)Sun, Yuan, Tsunoda, Vedral, Benjamin, and
  Endo]{sun2021mitigating}
Jinzhao Sun, Xiao Yuan, Takahiro Tsunoda, Vlatko Vedral, Simon~C Benjamin, and
  Suguru Endo.
\newblock Mitigating realistic noise in practical noisy intermediate-scale
  quantum devices.
\newblock \emph{\textit{Physical Review Applied}}, 15\penalty0 (3):\penalty0
  034026, 2021.

\bibitem[Tsamardinos et~al.(2006)Tsamardinos, Brown, and
  Aliferis]{max-min-hill-climbing}
Ioannis Tsamardinos, Laura~E. Brown, and Constantin~F. Aliferis.
\newblock {The max-min hill-climbing Bayesian network structure learning
  algorithm}.
\newblock \emph{\textit{Machine Learning}}, 65\penalty0 (1):\penalty0 31–78,
  2006.

\bibitem[Urbanek et~al.(2021)Urbanek, Nachman, Pascuzzi, He, Bauer, and
  de~Jong]{urbanek2021mitigating}
Miroslav Urbanek, Benjamin Nachman, Vincent~R Pascuzzi, Andre He, Christian~W
  Bauer, and Wibe~A de~Jong.
\newblock Mitigating depolarizing noise on quantum computers with
  noise-estimation circuits.
\newblock \emph{\textit{arXiv preprint arXiv:2103.08591}}, 2021.

\bibitem[Utkarsh et~al.(2020)Utkarsh, Behera, and Panigrahi]{behera2020solving}
Utkarsh, Bikash~K. Behera, and Prasanta~K. Panigrahi.
\newblock Solving vehicle routing problem using quantum approximate
  optimization algorithm.
\newblock \emph{\textit{arXiv preprint arXiv:2002.01351}}, 2020.

\bibitem[Verdon et~al.(2017)Verdon, Broughton, and Biamonte]{verdon2017quantum}
Guillaume Verdon, Michael Broughton, and Jacob Biamonte.
\newblock A quantum algorithm to train neural networks using low-depth
  circuits.
\newblock \emph{\textit{arXiv preprint arXiv:1712.05304}}, 2017.

\bibitem[Vovrosh et~al.(2021)Vovrosh, Khosla, Greenaway, Self, Kim, and
  Knolle]{vovrosh2021simple}
Joseph Vovrosh, Kiran~E Khosla, Sean Greenaway, Christopher Self, MS~Kim, and
  Johannes Knolle.
\newblock Simple mitigation of global depolarizing errors in quantum
  simulations.
\newblock \emph{Physical Review E}, 104\penalty0 (3):\penalty0 035309, 2021.

\end{thebibliography}
